\documentclass[11pt]{article}
\usepackage[utf8]{inputenc}
\usepackage{ulem}
\usepackage{subfigure}
\usepackage{comment}
\usepackage{multicol}
\usepackage{titlesec}
\usepackage{graphicx}
\usepackage{subcaption}
\usepackage{booktabs}
\usepackage{listings}
\usepackage[]{EMNLP2023}

\renewcommand{\footnotesize}{\tiny} 

\newcommand{\corp}{Google}
\title{CRQBench: A Benchmark of Code Reasoning Questions}
\author{Elizabeth Dinella \\
    Bryn Mawr College \\
  \texttt{edinella@brynmawr.edu} \\\And
  Satish Chandra \\
  Google, Inc. \\
  \texttt{schandra@acm.org} \\ \And
  Petros Maniatis  \\
  Google DeepMind \\
  \texttt{maniatis@google.com}
  }
  
\date{December 2023}
\begin{document}

\maketitle
\begin{abstract}
Large Language Models have demonstrated exceptional proficiency on coding tasks, but it is challenging to precisely evaluate their code reasoning ability. Existing benchmarks are insufficient as they are unrealistic and conflate semantic reasoning ability with performance on software engineering tasks. We introduce CRQBench, a benchmark of 100 C++ code reasoning questions and answers derived from contextualized code review comments. To curate CRQBench, we use an LLM assistant alongside human inspection, reducing manual effort. 
We conduct an evaluation of GPT-4 on CRQBench and find that it produces correct responses grounded in the given context for 65 of the 100 questions.
\end{abstract}

\section{Introduction}

Large Language Models (LLMs) have demonstrated  effectiveness in coding tasks and appear to understand deep semantic properties of code~\cite{codex, palm, llama}. However, evaluations across various tasks~\cite{swebench, robustAPI} show less promising results, suggesting that models may have a limited syntactic understanding of programs. To evaluate a model's semantic reasoning ability in isolation, a benchmark specifically tailored for code reasoning question answering is needed. 

The predominant benchmarks for evaluating LLMs trained on code are HumanEval~\cite{codex} and MBPP~\cite{MBPP}. They measure a model's ability to synthesize programs from docstrings. These text-to-code benchmarks are synthetic, handwritten, and involve generating a standalone function. Other, more realistic, benchmarks~\cite{swebench, robustAPI} are designed to evaluate code reasoning indirectly through a software engineering task, and as a result conflate the model's ability to perform reasoning with the ability to perform the downstream task. In this work, we set out to curate a real-world, contextualized, benchmark for evaluating semantic reasoning ability in isolation.



Ideally, a benchmark for evaluating semantic reasoning ability should reflect real-world programming scenarios. Code review comments present an appealing target for this as they are non-synthetic and tied to a surrounding code context. Through a study of contextualized code review comments at \corp{}
we find that a subset embody semantically deep questions about code, but a majority are superficial (related to refactoring or style). Furthermore, we find that comments are rarely concise and unambiguous questions. 
Although code review comments provide a source of authentic semantic queries, it is non-trivial to extract clean questions.

We present CRQBench: a benchmark of real-world, contextualized, \underline{c}ode \underline{r}easoning \underline{q}uestions. 
To reduce human curator effort, we propose a cooperative LLM and human-in-the-loop approach which leverages in-context learning~\cite{GPT3} to filter and rephrase code reasoning questions from code review comments. We reproduce our corporate results for open source release using Github pull request comments in the CodeReviewer dataset~\cite{codereviewer}.

In summary, our work presents a benchmark of 100 C++ (code reasoning questions, answer, code context) tuples derived from pull request comments in the CodeReviewer dataset. In addition, we present our curation technique as a re-usable methodology and evaluate its effectiveness in reducing manual effort in benchmark curation. Lastly, we evaluate GPT-4~\cite{gpt4} on CRQBench and find that it produces correct responses grounded in the given context for 65 of the 100 questions.




\section{Motivating Examples}
\label{sec:mot-ex}
In this section, we illustrate the presence of code reasoning questions in code review comments, while highlighting the challenges in extracting them. Reviewers' identities are anonymized.  \\
\textbf{Observation 1:} \uline{Most code review comments are not related to code reasoning}. Through a manual analysis, we find that a majority of Github (65\%) and \corp{} (80\%) code review comments  are not related to code reasoning. We consider a comment to be related to code reasoning if in order to ask, answer, or address, it requires reasoning 
over reachability, data flow, control flow, or program variable and state. 
Instead, code review comments are often shallow edit suggestions related to style, structure, documentation, or syntactic reasoning. Consider Figures~\ref{fig:ex3} (and~\ref{fig:ex4} in appendix), in which the reviewers make shallow comments regarding style and \textit{syntactic} reasoning respectively.
During our analysis, we also found comments that are discussions of the intended behavior or specification (Figure~\ref{fig:ex2} in appendix). 
We quantify the density of these comment categories in Table~\ref{tab:density}.
\begin{table}[h]
\begin{tabular}{r|cc}
\toprule
                               & \corp{} & Github  \\ \hline
CRQ        & 20\%                        & 35\%                         \\ 
Shallow Edit Suggestion        & 60\%                        & 35\%                         \\ 
Func Behavior Discussion & 20\%                        & 30\%                         \\ \bottomrule
\end{tabular}
\caption{Code Review Comments By Type.}
\vspace{-3mm}
\label{tab:density}
\end{table}

\textbf{Observation 2:}  \uline{Code review comments are often not phrased as questions.} Through our manual analysis, we find that even when the comment is related to code reasoning, it is very rarely phrased as a concise and unambiguous code reasoning question. 
Consider Figure~\ref{fig:ex1}, in which the comment is phrased as an edit suggestion (removing the call to \texttt{std::move}) rather than the underlying code reasoning question: \textit{Does calling \texttt{std::move} on the return value \texttt{s.releasePeerSet()} impact the program's behavior?}
Furthermore, the comment contains extraneous information, referencing another reviewer. In Figure~\ref{fig:ex5}, the comment is posed as a question, but it is overly verbose. It consists of two sentences, one of which is an extraneous edit suggestion related to functional behavior. The first sentence, although related to code reasoning, is ambiguous and not contextualized in the reviewed code. It does not explicitly state which program variables \textit{``something else''} encompasses. A concise, unambiguous rephrasing could be: \textit{Can \texttt{error\_code} hold a value other than \texttt{ECONNREFUSED} or \texttt{ECONNRESET}?}

We also observe that code reasoning questions can be categorized into two types of queries that encompass all CRQs: \texttt{VALUE} and \texttt{EQUIV} queries.
A \texttt{VALUE} query (Figure~\ref{fig:ex5}) asks about the value or possible value
of a variable or expression at a program point. An \texttt{EQUIV} (equivalence) query (Figure~\ref{fig:ex1}) asks if two segments of code have differences in behavior. \texttt{EQUIV} queries typically underlay an edit suggestion. 
We find that in both Github and \corp{} code review comments,
75\% of code reasoning questions are \texttt{EQUIV} queries while 25\% are \texttt{VALUE} queries.

\textbf{Observation 3:} \uline{Answers to rephrased questions are not readily available.}
During our manual analysis, we inspected the developer's responses to comments. Responses came in the form of a natural language reply and/or a code edit.
Answers in the form of a developer reply suffer from the same ambiguities and verbosity as the reviewer comments.
Answers in the form of an edit require careful manual inspection to connect the change to the underlying code reasoning question. Sometimes the comment is ignored and not addressed.



\begin{figure}
    \centering
    \includegraphics[width=\columnwidth]{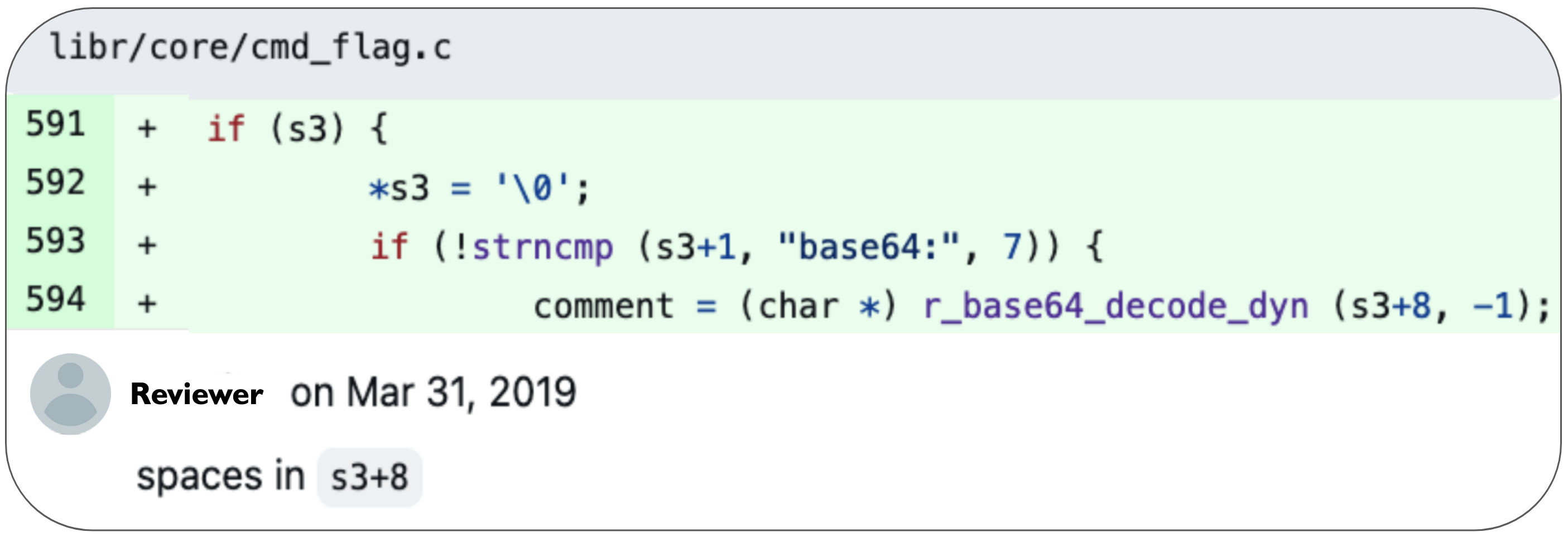}
     \caption{Shallow Edit Suggestion.\protect\footnotemark[2]} 
    \label{fig:ex3}

 \centering
    \includegraphics[width=\columnwidth]{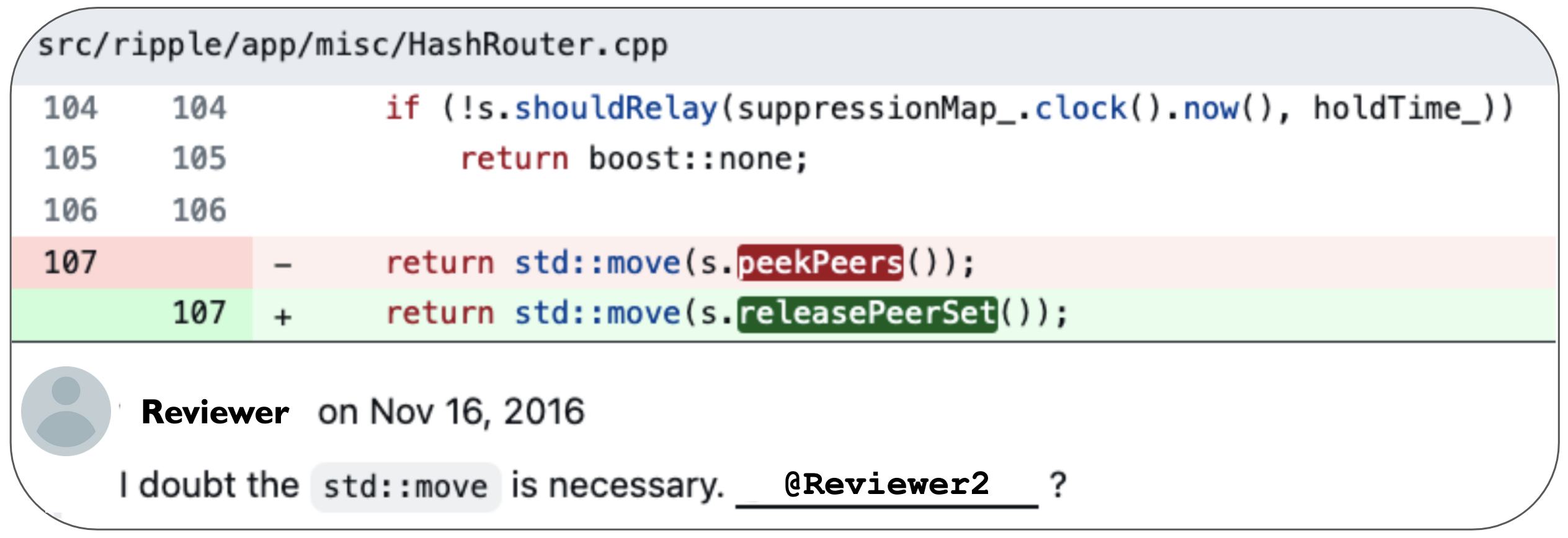}    
    \caption{Raw Code Review Comment.\protect\footnotemark[3]}
    \label{fig:ex1}

    \centering
    \includegraphics[width=\columnwidth]{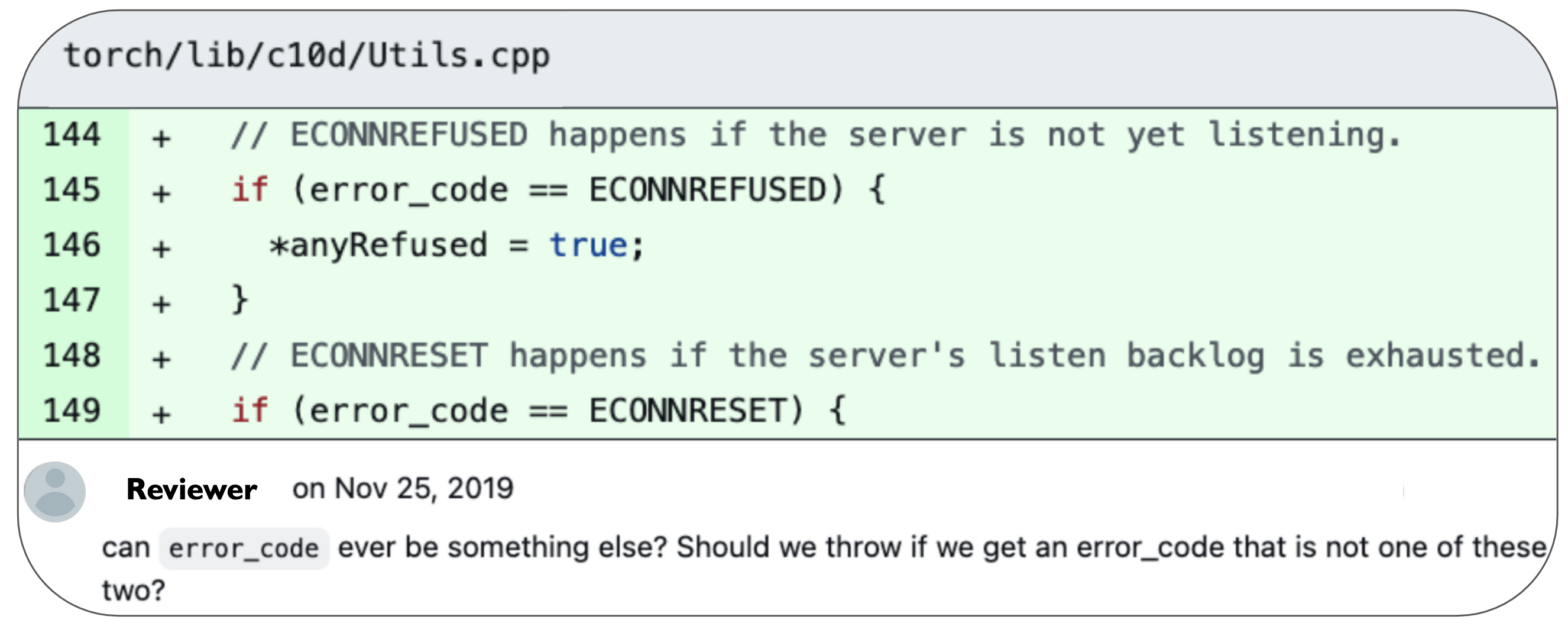} 
    \caption{Raw Code Review Comment.\protect\footnotemark[4]}
    \label{fig:ex5}
\vspace{-7mm}
\end{figure}


\section{Technique}
\begin{figure}
    \centering
    \includegraphics[width=\columnwidth]{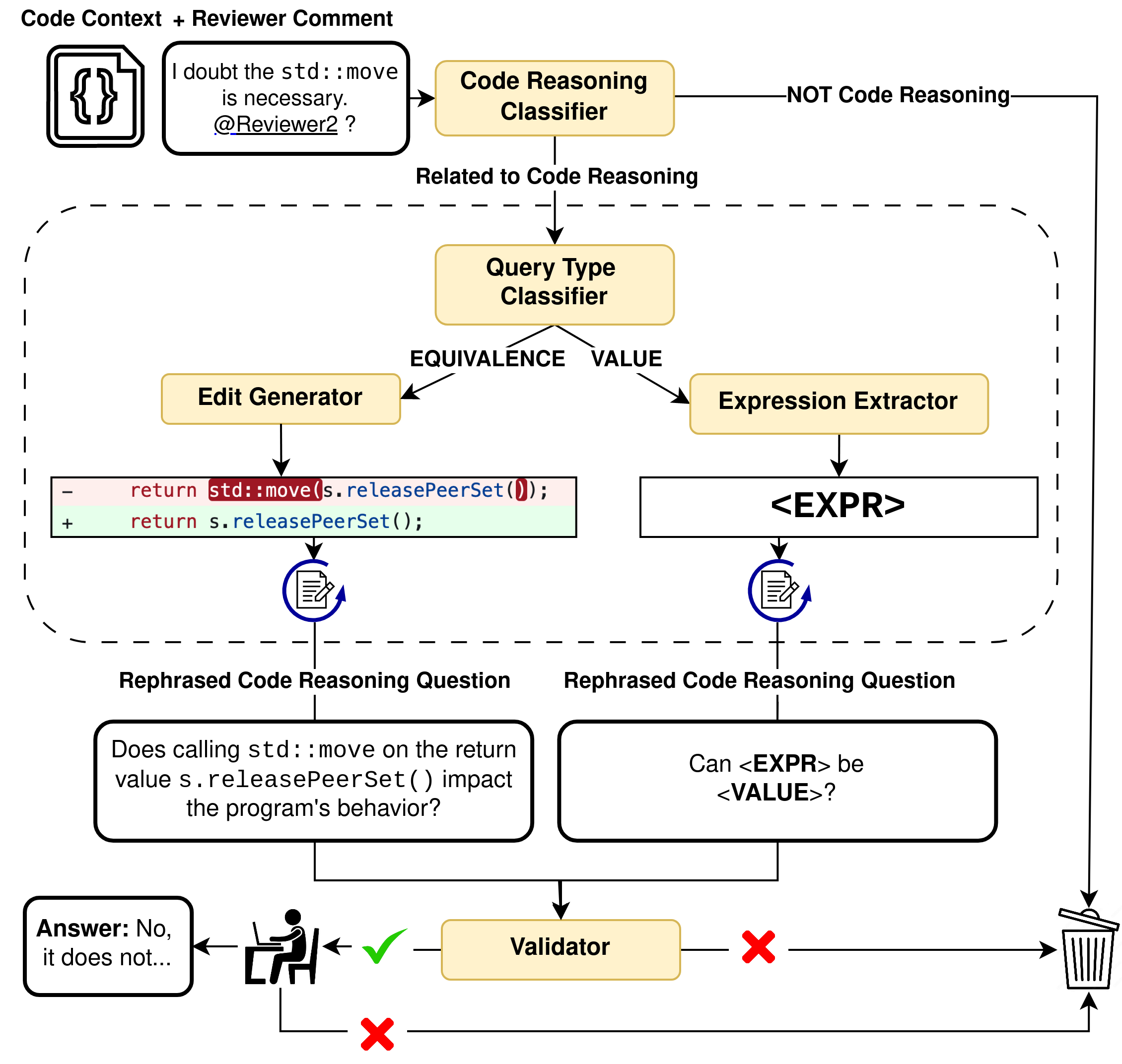}
    \caption{Benchmark Curation Methodology. Yellow boxes represent LLM prompts.}
    \label{fig:technique}
    \vspace{-5mm}
\end{figure}

Figure~\ref{fig:technique} illustrates our overall technique, which leverages a \corp{} code aware LLM in combination with human validation. 

\subsection{Classifying Comments} As discussed in \textbf{Observation 1}, a minority of code review comments are related to code reasoning. To reduce manual inspection, we create an LLM based \textsf{Code Reasoning Classifier} (Figure~\ref{fig:CodeReasoningClassifier} in appendix) which takes the raw reviewer comment and corresponding line of code and decides if it is related to code reasoning. 

We evaluate the performance of our \textsf{Code Reasoning Classifier} prompt on 100 randomly selected, manually labeled comments as shown in Table~\ref{tab:eval}.
We also experiment with a keyword matching approach using a hand derived list of undesirable keywords (Figure~\ref{fig:uninteresting_keywords} in appendix) \protect\footnotemark[5], but find it incurs significantly more false positives than our LLM classification. In summary, our classifier correctly identified 11 out of 20 \corp{} and 22 out of 35 GitHub code review comments as related to code reasoning, while misidentifying only 6 and 9 comments respectively.

\begin{table}[h]
\centering
\begin{tabular}{r|cc|cc}
\toprule
\small
& \multicolumn{2}{c|}{\corp{}}   & \multicolumn{2}{c}{Github}               \\ \hline
                & LLM  & KW & LLM  & KW  \\ \hline
Precision       & .64     & .31   & .71   & .39                  \\ 
Recall          & .52     & .81   & .63   & .1                  \\ 
F1 Score        & .57     & .45   & .67   & .56                   \\\bottomrule

\end{tabular}
\vspace{1mm}
\caption{Code Reasoning Classification performance of LLM and Keyword matching approaches.}
\label{tab:eval}
\vspace{-5mm}
\end{table}

\subsection{Rephrasing Comments as CRQs} As discussed in \textbf{Observation 2}, comments are rarely phrased as concise questions grounded over program elements. To avoid manual rephrasing, we again leverage the \corp{} LLM. This portion of our technique is shown in the dotted box in Figure~\ref{fig:technique} and is invoked on samples that have been accepted by the \textsc{Code Reasoning Classifier}. 
Our technique invokes different rephrasing techniques for each query type. The \textsf{Query Type Classifier} (Figure~\ref{fig:QueryTypeClassifier} in appendix) classifies a comment as either an \texttt{EQUIV} query or \texttt{VALUE} query, triggering the appropriate rephrasing technique based on the classification.
\\ 
\\
When the \textsf{Query Type Classifier} decides the comment is an \texttt{EQUIV} query, we use chain of thought~\cite{CoT} reasoning to effectively rephrase. Since \texttt{EQUIV} queries are typically underlying edit suggestions, we employ an LLM based \textsf{Edit Generator} (Figure~\ref{fig:edit-generator} in appendix) to perform the reviewer's suggested edit as a link in a chain of thought. The edit is leveraged to rephrase the reviewer comment using a few shot prompt (Figure~\ref{fig:equiv-rewriter} in appendix). 


When the \textsf{Query Type Classifier} decides the comment is a \texttt{VALUE} query, we similarly use a two step inference process similar to a chain of thought.
As a first step, our \textsf{Expression Extractor} uses a few-shot prompt (Figure~\ref{fig:expr-extractor} in appendix) leveraging the code context and reviewer comment to extract the relevant program expression: \texttt{<EXPR>}.
The relevant \texttt{<EXPR>} is used as a link to rephrase the reviewer comment  as a code reasoning question over the given expression using a few-shot prompt (Figure~\ref{fig:value-rewriter} in appendix).
\\
\\
Lastly, the rephrased question is given to the \textsf{Validator} for a self-consistency~\cite{selfconsistency} check to reduce the occurrence of poorly rephrased code reasoning questions that are not faithful to the original line comment. The \textsf{Validator} prompt (Figure~\ref{fig:Validator} in appendix) asks the LLM to decide if, given the original code, the reviewer comment has the same meaning as the rephrased comment. If the LLM \textsf{Validator} confirms the consistency, the rephrased question is selected as a confident candidate, and given to a human inspector to validate. 

We evaluate our technique's effectiveness in rephrasing code review comments into concise and unambiguous code reasoning questions. Our rephrasing approach (entire dotted box component in Figure~\ref{fig:technique})
is evaluated on both \corp{} (150 samples) and Github (160 samples) code review comments
that were flagged as related to code reasoning by our \textsf{Code Reasoning Classifier}. The samples were manually inspected and labelled as correct if they were concise, unambiguous, and faithful to the original reviewer comment. We achieved a precision of 66 on \corp{} code review comments and .63 on Github pull request comments.

\subsection{Evaluation}
We evaluate our methodology in terms of manual human curation required. In a purely manual approach, a human curator would need to inspect and classify 500 \corp{} (or 285 Github) code review comments and manually rephrase 100 questions. Using our proposed methodology, a human curator would need to inspect only 150 \corp{} (or 160 Github) code review comments without the need for any manual rewriting. 
Figure~\ref{fig:funnelgoogle} (in appendix) and Figure \ref{fig:funnelgithub}, respectively,
illustrate this comparison. The pencil indicates a manual rephrasing while the magnifying glass indicates inspection using our proposed technique. In summary, our cooperative LLM + human validation approach reduces the number of samples required to inspect by 1.8x on Github pull request comments and 3.3x on \corp{} code review comments.

\begin{figure}
    \centering
    \includegraphics[width=.9\columnwidth]{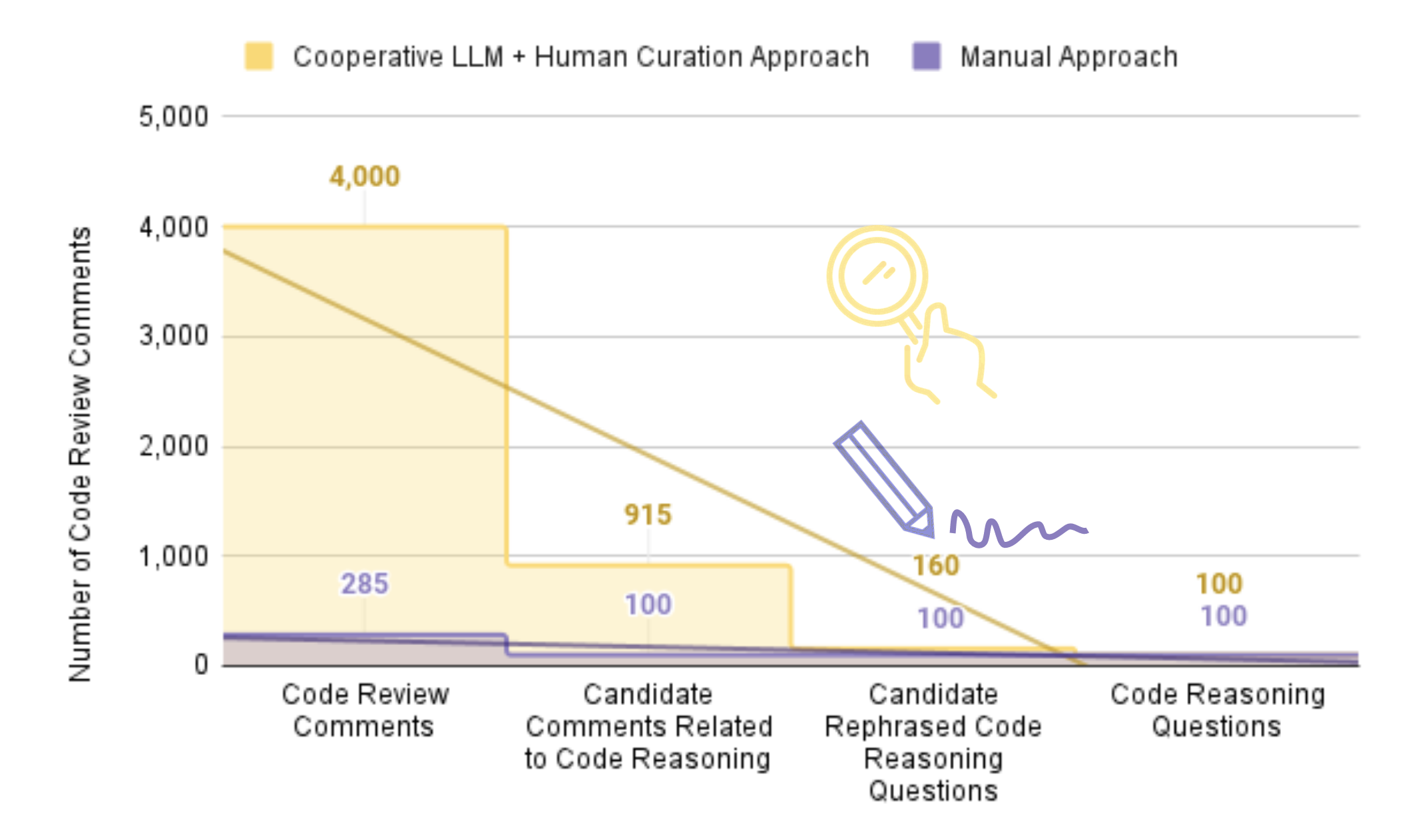}
    \caption{Manual vs Cooperative LLM Curation on Github Pull Request Comments.}
    \label{fig:funnelgithub}
    \vspace{-5mm}
\end{figure}

\section{GPT-4 Performance}
We evaluate GPT-4 on our benchmark by prompting it with the surrounding function context (Figure~\ref{fig:gpt-prompt} in Appendix). We evaluate the outputs manually considering a response to be accurate if it is both correct and contextually relevant. Correctness refers to the technical accuracy of the generated natural language response. Contextual relevance refers to the degree that the response is grounded in the given code context. We find the GPT-4 provides an accurate response on 65 of the 100 queries and are almost always (94\%) grounded in the given code context.


\begin{table}[h]
\centering
\begin{tabular}{r|ccc}
\toprule
\small
                    & Acc     & Total    & \% \\ \hline
                    & 65      & 100   & 65\%                  \\ 
\hline
\texttt{VALUE}      & 33     & 54   & 61\%                \\ 
\texttt{EQUIV}      & 32     & 46   & 70\%                    \\\bottomrule

\end{tabular}
\vspace{1mm}
\caption{Performance of GPT-4 on CRQBench.}
\label{tab:eval}
\vspace{-3mm}
\end{table}

Lastly, we conducted an error analysis to categorize the 35 incorrect responses. The majority of errors (25 instances) were due to the model lacking necessary context, such as usages of the given function, definitions of a used function or macro, or usages of a variable. Five errors were attributed to gaps in C++ knowledge, and the remaining five were due to incorrect evaluation of logic. Examples of each error scenario are shown in the Appendix (Figures~\ref{fig:gpt_err_logic} - \ref{fig:gpt_err_cpp}). 

We also experimented with evaluating the 7 billion parameter open source model Falcon~\cite{falcon}, but found it to have a much lower accuracy (\~25\%) as it is a significantly smaller model. 
\section{Limitations}

\subsection{Extracting Answers to CRQs}  To extract answers, we use an entirely manual based approach. A human curated an answer through a best effort approach by  inspecting the cloned repository at the commit being reviewed. The answer is derived by reasoning over the code context, edit made (or not), and developer textual responses in the comment thread. In essence, our benchmark gathers the response which was implicitly provided by the developer, rather than an answer verified by a symbolic program analysis approach. We default to manual curation of answers due to the challenges presented in Section~\ref{sec:mot-ex}. 

\subsection{Size of Target Environment}
Although the number of samples to inspect or rephrase is greatly diminished with our approach, the total number of comments needed to arrive at 100 code reasoning questions is much larger. Our cooperative approach requires greater than 10x more code review comments to derive 100 CRQs. This is due to false negatives in \textsf{Code Reasoning Classifier} and the \textsf{Validator}. 


\footnotetext[2]{https://github.com/radareorg/radare2/pull/13555\#discussion\_r270676564}
\footnotetext[3]{https://github.com/XRPLF/rippled/pull/1904\#discussion\_r88226072}
\footnotetext[4]{https://github.com/pytorch/pytorch/pull/30354\#\\discussion\_r350485189}
\footnotetext[5]{We also experimented with a hand derived desirable keyword list (Figure~\ref{fig:interesting_keywords} in appendix). We report results for the undesirable keyword approach as it achieved a higher F1 score.}

\bibliographystyle{ACM-Reference-Format}
\bibliography{refs}

\pagebreak
\onecolumn

\appendix
\section{Appendix}
\lstset{
    basicstyle=\ttfamily,
    breaklines=true, 
    xleftmargin=\parindent,
    xrightmargin=\parindent,
    basicstyle=\small\ttfamily
}
\begin{figure}[h]

\begin{minipage}{.5\columnwidth}
\centering
\includegraphics[width=\columnwidth]{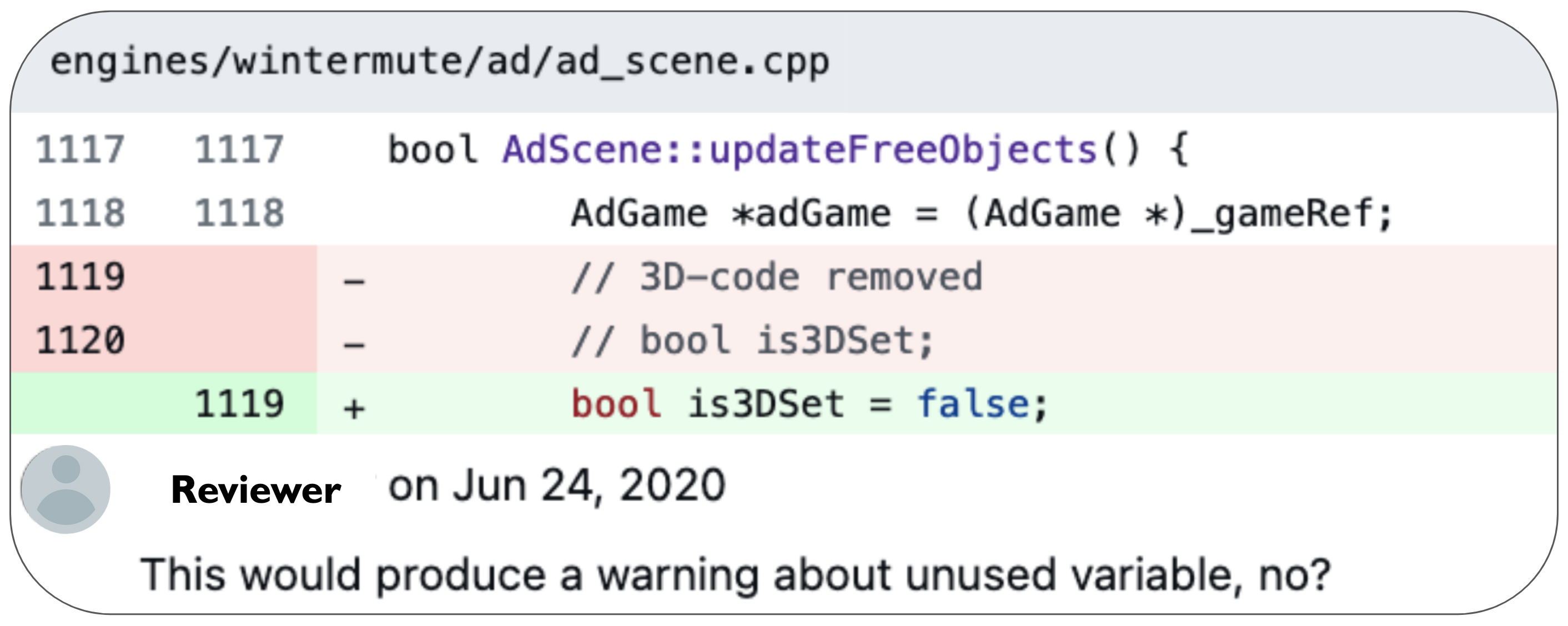}
    \caption{Shallow Edit Suggestion.\protect\footnotemark[5]} 
    \label{fig:ex4}
\end{minipage}

\end{figure}

\begin{figure}[h]
    \begin{minipage}{.5\columnwidth}
\centering
\includegraphics[width=\columnwidth]{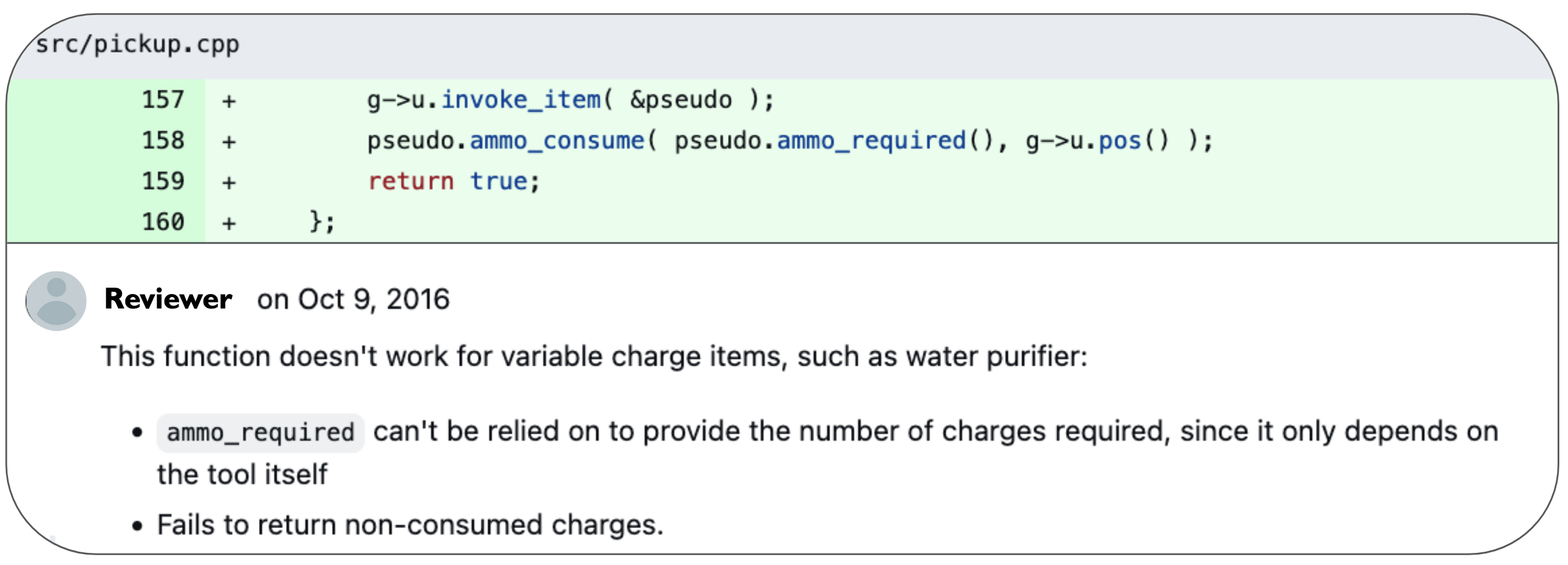}
    \caption{Functional Behavior Discussion.\protect\footnotemark[6]}
    \label{fig:ex2}
\end{minipage}
\end{figure}
\footnotetext[5]{https://github.com/scummvm/scummvm/pull/2347\#discussion\_r445103384}
\footnotetext[6]{https://github.com/CleverRaven/Cataclysm-DDA/pull/18691\#discussion\_r82533193}

\begin{figure}[h]
    \begin{minipage}{\columnwidth}
\centering
\includegraphics[width=\columnwidth]{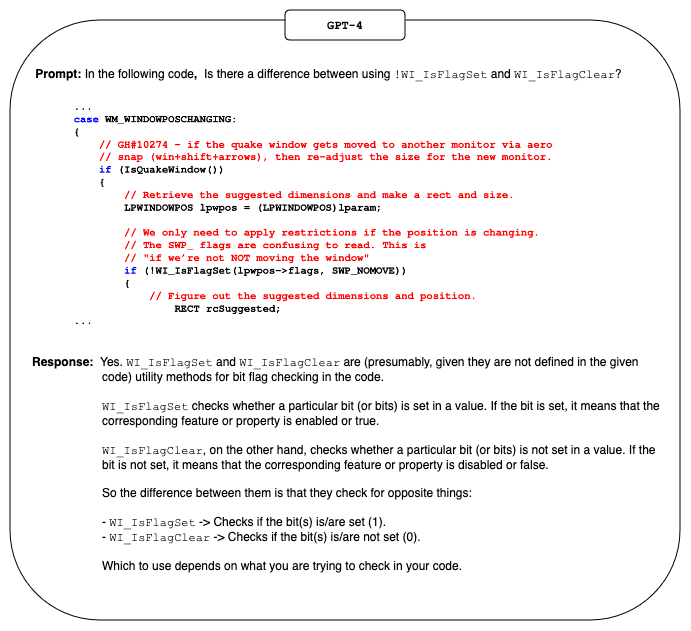}
    \caption{Incorrect GPT Response: Logic Error. The model recognized that the flags check for opposite conditions but failed to deduce that the inverse of one flag is equivalent to the other.}  
    \label{fig:gpt_err_logic}
\end{minipage}
\end{figure}

\begin{figure}[h]
    \begin{minipage}{\columnwidth}
\centering
\includegraphics[width=\columnwidth]{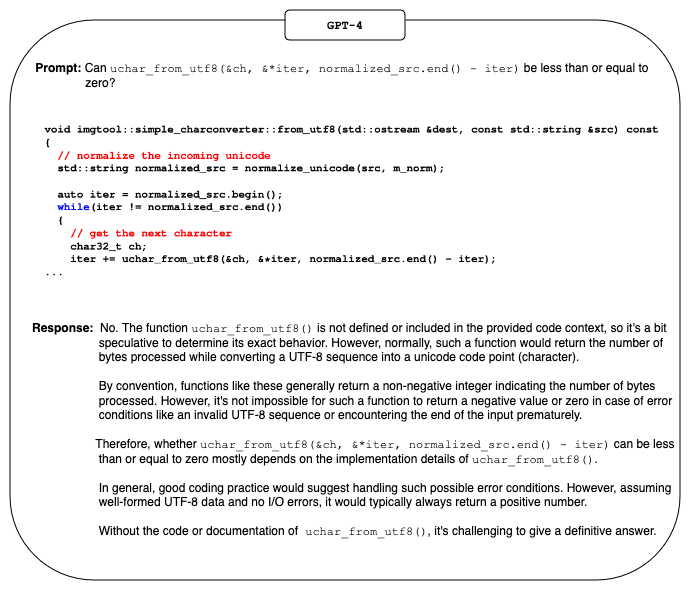}
    \caption{Incorrect GPT Response: Missing Context. The model was unable to determine if the expression could be less than or equal to zero as it did not have access to the function definition.}
    \label{fig:gpt_err_missing_ctx}
\end{minipage}
\end{figure}

\begin{figure}[h]
    \begin{minipage}{\columnwidth}
\centering
\includegraphics[width=\columnwidth]{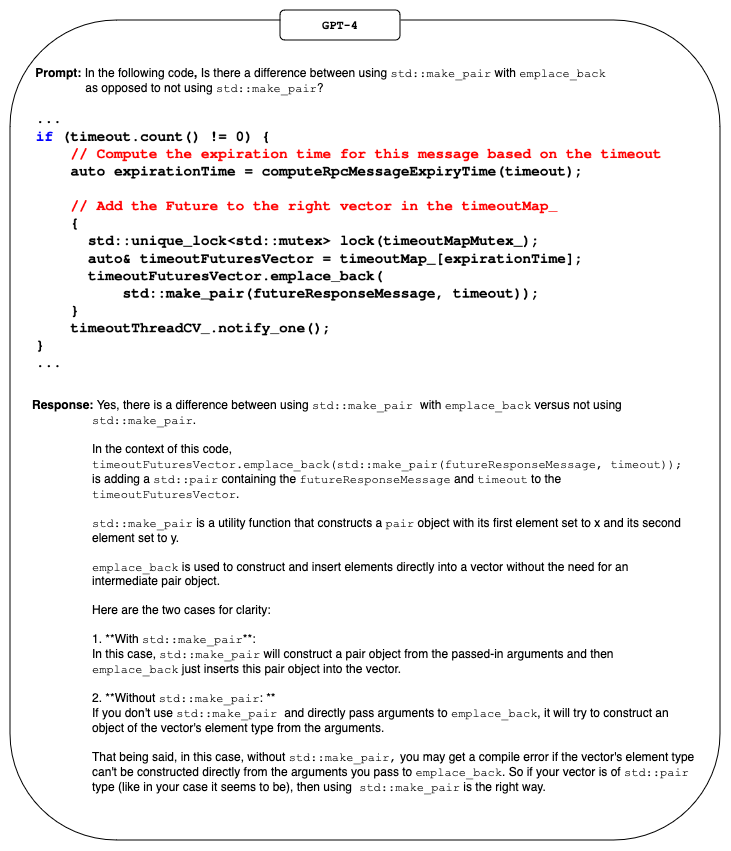}
    \caption{Incorrect GPT Response: Incorrect C++ knowledge. The model incorrectly responded Yes. In particular, the model was incorrect that without using \texttt{std::make\_pair}, a compile error may occur.}
    \label{fig:gpt_err_cpp}
\end{minipage}
\end{figure}

\begin{figure}[h]
\begin{lstlisting}[]
Given the following code, {QUESTION}
{SURROUNDING_FUNCTION}
\end{lstlisting}
    \caption{Prompt to GPT-4 for CRQBench evaluation.}
        \label{fig:gpt-prompt}
\end{figure}

\begin{figure}[h]
\begin{lstlisting}[]
You are a senior, expert C/C++ programmer with a lot of experience analyzing C/C++ code.
Given a C/C++ line comment:

{COMMENT}

associated with this line of code:

```
{CODE_SEGMENT}
```

please determine which of the following categories best classifies that
line comment:

1. unknown: unknown
2. reasoning: a code reasoning question such as requiring control-flow or value propagation
3. explanation: a question asking what the code does
4. structure: request to reorganize or refactor code, such as extracting classes or methods
5. planning: prioritization, planning, or choosing what to work on next
6. style: a code style, or readability question including things like renaming symbols
7. format: a comment or question about code formatting

and provide an explanation of why that comment should be in that specific category.
If you are uncertain about the category, respond with "0".

Generate your output in YAML format like this:

Explanation: <why you chose a specific category>
Line comment category: <category>

Response:
```yaml
\end{lstlisting}
    \caption{\texttt{Code Reasoning Classifier} Prompt.}
        \label{fig:CodeReasoningClassifier}
\end{figure}

\begin{figure}
\begin{lstlisting}[]
You are an expert software engineer, asked to determine whether or not a reviewer comment is about substituting code.

Here are some examples.

Example 1:

LINE_COMMENT:
```
also check if the tf example is empty?
```

1. No.

Example 2:

LINE_COMMENT:
```
This stores a pointer to a temporary object. This pointer becomes invalid right after this statement. I suggest you initialise it with a `nullptr` and check for that.
'''

1. No.

Example 3:

LINE_COMMENT:
```
Please use stolen(fduri) instead of hardcoding 5
'''

1. Yes.

Example 4:

LINE_COMMENT:
```
The copy of a single int (previous type of fg, bg) is fine here, however copying the vector is not good. Use a const reference instead.
'''

1. Yes.

Example 5:

LINE_COMMENT:
```
{COMMENT}
'''

1.
\end{lstlisting}
    \caption{\texttt{Query Type Classifier} Prompt.\protect\footnotemark[7]}
    \label{fig:QueryTypeClassifier}
\end{figure}

\footnotetext[7]{The prompt used in our technique uses \corp{} internal code review comments. To share the prompt, comments are substituted with open source code reviews from Github. We attempt to find substitutions which are similar to our \corp{} code based prompt.}

\begin{figure}
\begin{lstlisting}[]
You are an expert C++ coder and know how to review C++ code at a Corporation, and how to 
respond to reviewer comments. 
In the code below, a reviewer has left a comment marked by LINE_COMMENT.

Code:
```
{CODE_SEGMENT_WITH_COMMENT}
'''

The comment is asking to consider the following original code and modified code:
\end{lstlisting}
    \caption{\texttt{Edit Generator} Prompt.}
    \label{fig:edit-generator}
\end{figure}

\begin{figure}
\begin{lstlisting}[]
You are an expert C++ coder and know how to review C++ code at a Corporation, and how to respond to reviewer comments. In the code below, a reviewer has left a comment marked by LINE_COMMENT.

Please answer the following question.
1. Can you please list the relevant program expression?

Here are some examples.
Example 1:

Code:
```
int comfort = 0;
comfort_response_t comfort_response;
comfort_response.aid = &item( "null" );
[*] [LINE_COMMENT] This stores a pointer to a temporary object. This pointer becomes invalid right after this statement. I suggest you initialise it with a nullptr and check for that.
bool plantsleep = has_trait( trait_CHLOROMORPH );
bool fungaloid_cosplay = has_trait( trait_M_SKIN3 );
bool websleep = has_trait( trait_WEB_WALKER );
```

LINE_COMMENT: This stores a pointer to a temporary object. This pointer becomes invalid right after this statement. I suggest you initialise it with a nullptr and check for that.

1. comfort_response.aid

Example 2:

Code:
```
Value *Arg = Call->getArgOperand(0);
Value *LoweredArg = getLoweredByValOperand(Arg, Builder);
for (Value *A : Call->arg_operands()) {
    DXASSERT(A == Arg, "oops");
[*] [LINE_COMMENT] how could A be different unless there's a fundamental problem with operand iteration or something?
}
HLModule::MarkPreciseAttributeOnValWithFunctionCall(LoweredArg, Builder, *m_pModule);
addToDeadInsts(Call);
'''

LINE_COMMENT: how could A be different unless there's a fundamental problem with operand iteration or something?

1. A

Example 3:

Code:
```
{CODE_SEGMENT_WITH_COMMENT}
'''

LINE_COMMENT: {COMMENT}
\end{lstlisting}
    \caption{\texttt{Expression Extractor} Prompt.\protect\footnotemark[7]}
    \label{fig:expr-extractor}
\end{figure}


\begin{figure}
\begin{lstlisting}[]
Can you please rephrase a reviewer LINE_COMMENT that suggests the following code change? The rephrased comment should include program elements. Here are a few examples.

Example 1:
Before:
```
void gyroDataAnalyse(const gyroDev_t *gyroDev, biquadFilter_t *notchFilterDyn) {
    static FAST_RAM float fftAcc[XYZ_AXIS_COUNT] = {0, 0, 0};
    static FAST_RAM uint32_t fftAccCount = 0;
```
After:
```
void gyroDataAnalyse(const gyroDev_t *gyroDev, biquadFilter_t *notchFilterDyn) {
    static FAST_RAM float fftAcc[XYZ_AXIS_COUNT];
    static FAST_RAM uint32_t fftAccCount = 0;
```
[*] [LINE_COMMENT] Zero-initialized by default as well.
Rephrased: Is there a difference between initializing fftAcc to {0, 0, 0} and not initializing fftAcc?

Example 2:
Before:
```
if (isIntegralType(iter.dtype(), false)) {
    AT_DISPATCH_INTEGRAL_TYPES(iter.dtype(), "fmod_cpu", [&]() {
        cpu_kernel(iter, [](scalar_t a, scalar_t b) -> scalar_t {
            return std::fmod(a, b);
        });
    });
 }
```
After:
```
 if (isIntegralType(iter.dtype(), false)) {
    AT_DISPATCH_INTEGRAL_TYPES(iter.dtype(), "fmod_cpu", [&]() {
        cpu_kernel(iter, [](scalar_t a, scalar_t b) -> scalar_t {
            return a % b;
        });
    });
 } 
```
[*] [LINE_COMMENT] Should we use fmod or % here?
Rephrased: Is there a difference between computing the modulus of a and b using std::fmod vs using the binary operator %?

Example 3:
Before:
```
{BEFORE}
```
After:
```
{AFTER}
```
[*] [LINE_COMMENT] {COMMENT}
\end{lstlisting}
    \caption{\texttt{Equiv Rewriter} Prompt.\protect\footnotemark[7]}
    \label{fig:equiv-rewriter}
\end{figure}

\begin{figure}
\begin{lstlisting}[]
You are an expert C++ coder and know how to review C++ code at a Corporation, and how to respond to reviewer comments.
A reviewer has left a comment marked by LINE_COMMENT.

Can you please rephrase the LINE_COMMENT as a question over program expressions?
The Question should start with the word "Can".

Here are some examples.

Example 1:

Comment:
```
This stores a pointer to a temporary object. This pointer becomes invalid right after this statement. 
I suggest you initialise it with a nullptr and check for that.
```

Program Expression: comfort_response.aid
Question: Is there a difference between initializing comfort_response.aid with &item( "null" ) as opposed to nullptr?

Example 2:

Comment:
```
how could A be different unless there's a fundamental problem with operand iteration or something?
```

Program Expression: A
Question: Can A ever be equal to Arg?

Example 3:

Comment:
```
{COMMENT}
```

Program Expression: {EXPR}
\end{lstlisting}
    \caption{\texttt{Value Rewriter} Prompt.\protect\footnotemark[7]}
    \label{fig:value-rewriter}
\end{figure}

\begin{figure}
\begin{lstlisting}[]
You are an expert C++ coder and know how to review C++ code at a Corporation, and how to respond to reviewer comments.
In the code below, a reviewer has left a comment marked by LINE_COMMENT.

Code:
```
{CODE}
```

Given the LINE_COMMENT, answer the following question with YES or NO:
Does ```{Q_C}``` have the same meaning as the LINE_COMMENT?
\end{lstlisting}
    \caption{\texttt{Validator} Prompt.}
    \label{fig:Validator}
\end{figure}

\begin{figure}\begin{verbatim}
Interesting Keywords:
"except" 
"segfault"
" fault"
"precondition"
 "assumption"
"undefined behavior"
 " ub " 
"null", 
"reach"
"ever be true"
"ever be false"
"branch taken"
"branch not taken"
"deref"
"reference"\end{verbatim}
    \caption{Desirable Keywords for Positive Matching.}
    \label{fig:interesting_keywords}
\end{figure}
\begin{figure}\begin{verbatim}
Uninteresting Keywords:
"test"
“nit” 
“follow up”
“log a higher level” 
"log a lower level"
"logging"
“naming”
“readability”
“TODO”
“description”
“comment”
"typo"
"clang"
"style guide"
“period"
"restructure"
"restructuring"
"refactor"
"move"
"offline"
"space"
"spacing"\end{verbatim}
\caption{Undesirable Keywords for Negative Matching.}
\label{fig:uninteresting_keywords}
\end{figure}

\begin{figure}
    \centering
    \includegraphics[width=0.5\textwidth]{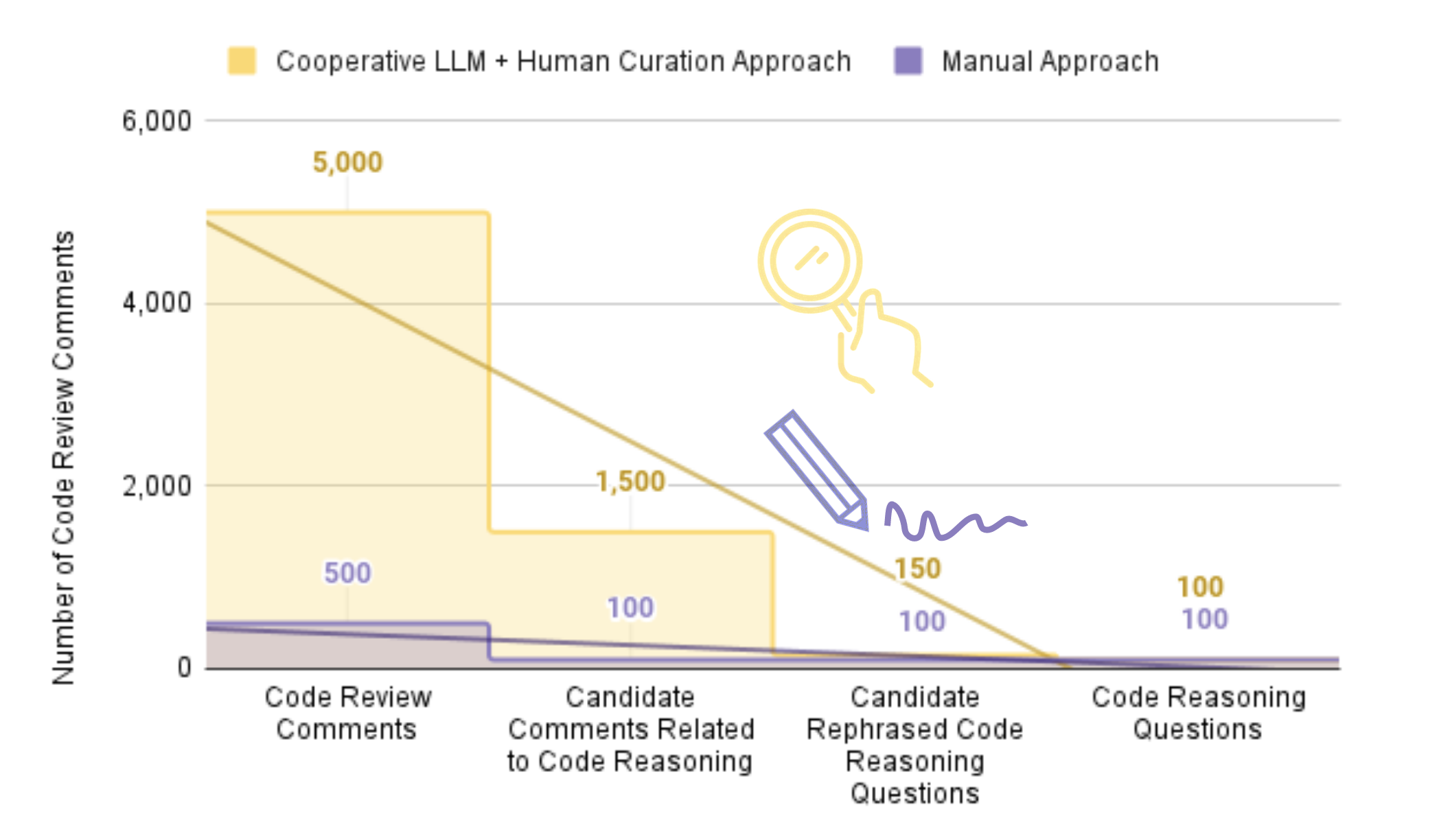}
    \caption{Manual vs Cooperative LLM Curation approaches on \corp{} Code Review Comments.}
    \label{fig:funnelgoogle}
\end{figure}
\end{document}